# Linear Predictive Coding as an Estimator of Volatility

## Louis Mello

**Abstract** In this paper, we present a method of estimating the volatility of a signal that displays stochastic noise (such as a risky asset traded on an open market) utilizing Linear Predictive Coding. The main purpose is to associate volatility with a series of statistical properties that can lead us, through further investigation, toward a better understanding of structural volatility as well as to improve the quality of our current estimates.

Let us take, as a point of departure, Shannon's definition of the uncertainty of probabilities associated to a given distribution:

$$H(p) = -\sum_i p_i \log p_i \quad (1.1)$$

i.e. information entropy.

We may define *Maximum Entropy* as the precept that permits us to formulate a model whose bias is reduced as the result of maximizing $H(p)$. This should elucidate the concept that (1.1) is a close relative of Gibbs' characterization, in statistical mechanics, of the technique of minimizing the average log probability,

$$H = \sum_i p_i \log p_i \quad (1.2)$$

subject to constraints in the form of expectation values, in order to determine the best probability distribution of a given open canonical system. That means that the amount of information, or uncertainty, output by an information source is a measure of its entropy. In turn, the entropy of that very source determines the amount of bits per symbol required to encode the source's information.

*Maximum Entropy* relies on the work of Information Theory as well as on the tenets of Bayesian statistics in that it makes use of epistemic probabilities, i.e., the **explicit use of prior information**. This characterizes the process described below,

Linear Predictive Coding, insofar as it is theoretically accurate to assume that the conditional variance of processes that exhibit dispersion, or volatility, display weak persistence (short-term memory) to varying degrees[1]. It has been shown that ARCH/GARCH (Autoregressive Conditional Heteroskedastic) processes reveal serial correlation for some period, after which they recover a simple uncorrelated Itô process.

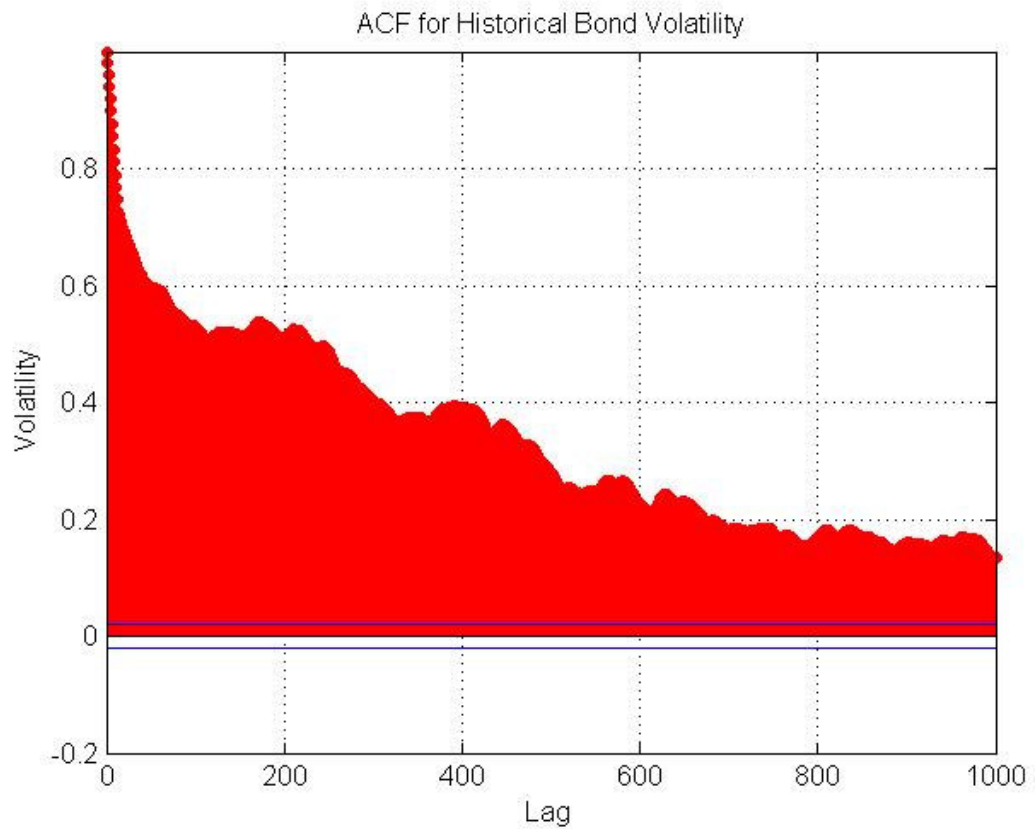

As can be seen from the graph above, historical volatility, defined as:

$$v_i = \frac{1}{n} \sum_{j=i}^{i+n-1} \sqrt{\left(\bar{x}_i - x_j\right)^2} \qquad (1.3)$$

is no more than an *n* period sliding standard deviation. The resulting process displays significant long-term autocorrelation, suggesting fractional Brownian motion (fBm).

---

[1] **Engle, Robert --** *Autoregressive Conditional Heteroskedasticity with Estimates of the Variance of U.K. Inflation*, **Econometrica,** 50 (1982): 987-1008.

## Linear Prediction

The **Linear Predictive Coding** (LPC) concept, which originated in the study of speech synthesis, is based on an attempt to model a signal where a point, $x_n$, of the data set bears correlation to the $P$ previous ones through the equation:

$$x_n = \sum_{i=1}^{P} af_i \cdot x_{n-1} \qquad (1.4)$$

or from the $P$ following ones :

$$x_n = \sum_{i=1}^{P} ab_i \cdot x_{n+i} \qquad (1.5)$$

called the Linear Prediction equations. The $af_i$ and the $ab_i$ are called, respectively, the forward prediction coefficients and the backward prediction coefficients (they have also been labeled the forward and backward autoregressive coefficients).

In point of fact, in the world of finance there is no such thing as a signal without noise. Thus, equations (1.4) and (1.5) are only approximations, and should be expressed as:

$$x_n = \sum_{i=1}^{P} af_i \cdot x_{n-i} + sf_n \qquad (1.6)$$

and

$$x_n = \sum_{i=1}^{P} ab_i \cdot x_{n-i} + sb_n \qquad (1.7).$$

These equations, in turn, can be rewritten as:

$$sf_n = x_n - \sum_{i=1}^{P} af_i \cdot x_{n-i} \qquad (1.8)$$

and

$$sb_n = x_n - \sum_{i=1}^{P} ab_i \cdot x_{n+i} \qquad (1.9)$$

It is obvious that what is predicted in (1.8) and (1.9) is the noise, and we use the fact that the noise is not correlated to the signal in order to uncover the autocorrelation structure of the original data set.

### *The Prediction Coefficients*

There are several ways to obtain the values of the prediction coefficients from the data set, all of which are based on the fact that equations (1.3) and (1.4) can be seen as a matrix relation:

$$\mathbf{X} = \mathbf{Y} \times \mathbf{A} \qquad (1.10)$$

Where:

$\mathbf{X} = x_n, \mathbf{Y} = y_{k_1} = x_{k_{-1}}$ and $\mathbf{A} = af_i$

and

$\mathbf{Y} = P \times (N - P)$ matrix.

In this matrix relation, $\mathbf{X}$ and $\mathbf{Y}$ can be constructed from the data set and $\mathbf{A}$ is unknown. In order to find the values of $af_i$ this linear system must be inverted.

There are a few observations regarding this matrix:

1. The matrix **Y** has $P$ rows and $(N - P)$ columns, where $N$ is the number of total data points;
2. The system need not be symmetric; there is no reason to have the number of data points in one-to-one correspondence to the number of signals;
3. One cannot extract more than $f(N;2)$ sinusoids from a signal of length $N$;
4. Usually $N$ is large compared to $P$, so the linear system is highly redundant;
5. The matrix **Y** has a specific form, constructed from a single vector, and as such is also redundant: It is, clearly, a Toeplitz matrix;
6. The Toeplitz form of the matrix **Y** should help the inversion step.

In effect, what we are attempting is the use of $P$ consecutive values from our original signal to predict the value in $P_{i+1}$. The model assumes stationarity; that is, the existence of serial correlation uniquely with respect to the absolute differences in the indices of the signal (time domain). The proof of the Burg method, which is the form we have chosen to implement, uses a maximization of the entropy quantity in the time domain, and as such, has been called the Maximum Entropy or MEM method by Burg himself.

## The Burg Algorithm

The Burg algorithm is a form of forward-backward prediction model. Based on the Levinson recursion, the following recursive computation has been derived:

$$\begin{cases} ef_P(n) = ef_{P-1}(n) + k_P \cdot eb_{P-1}(n-1) \\ eb_P(n) = eb_{P-1}(n-1) + k_P \cdot ef_{P-1}(n) \end{cases} \quad (1.11)$$

And the coefficients are computed with this expression:

$$\hat{k}_m = \frac{2 \sum_{a=m+1}^{N} e_{m-1}^f(n) e_{m-1}^b(n-l)}{\sum_{a=m+1}^{N} \left| e_{m-1}^f(n) \right|^2 + \sum_{a=m+1}^{N} \left| e_{m-1}^b(n-l) \right|^2} \quad (1.12)$$

The figure below shows a clear diagram of the recursive computation of Burg algorithm.

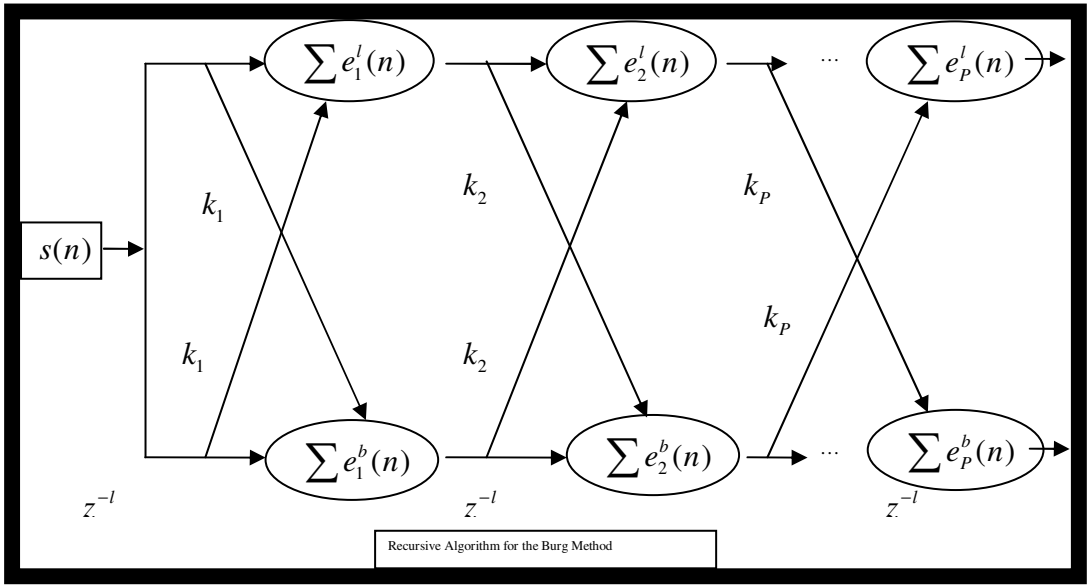

Recursive Algorithm for the Burg Method

The Burg method uses the last three points (plus some others) to execute the inversion of the matrix. This process is meant to be applied on signals for which the forward and the backward prediction coefficients are equal.

The algorithm is very fast, usually significantly faster than a Fourier transform, and performs very well on large data sets (processing time is $O(NxP)$).

*Prediction*

Now, if we define the set of coefficients obtained from (1.12) as $d_j$, our estimation equation will be given simply by:

$$\widehat{y}_n = \sum_{j=1}^{M} d_j y_{n-j} + x_n \qquad (1.13)$$

We apply this equation to our data set in order to ascertain the magnitude of the discrepancies $x_i$. In our model, the mean square discrepancy is reintroduced into the system as an offset to the initial sum in (1.13).

It is noteworthy that (1.13) is not a simple extrapolation process, although it does represent the special case of a linear filter; it is more powerful and complex than either a straight line or a low-order polynomial.

Also, in spite of the noise not being directly part of the equation, it is taken into consideration as long there is no pair wise correlation. The method actually estimates the diagonal of the matrix.

The literature abounds with suggestions on how one could increase the stability of the model by redirecting all roots of the characteristic polynomial back into the unit circle. We have decided not to adopt such measures for two main reasons:

1. The linear prediction is short term, and the instabilities increase at a very slow pace;

2. By processing both a forward and backward extrapolation, we can see that there is a reasonably good agreement between the two; hence, there is no need to consider instability an obstacle to the accuracy of the model.

## Results

By way of comparison, we have measured LPC against the GARCH (1,1) model. Both are autoregressive in nature; however, the coefficients for the LPC method can be computed without the need for an approximate parameter search via optimization, and the number of future points can be $> t + 1$, which is not the case for GARCH. We also note that LPC is more robust than GARCH, inasmuch as it does not assume any particular probability distribution.

The table below illustrates the estimates:

| Asset | LPC | GARCH |
|---|---|---|
| 10 Yr. Bond | 0.09760 | 0.09998 |
| Bovespa | 0.43770 | 0.44924 |
| Dow | 0.09200 | 0.09050 |
| Google | 0.39963 | 0.36894 |
| MSFT | 0.15222 | 0.14189 |
| Nikkei | 0.27408 | 0.28060 |
| S&P | 0.10029 | 0.09138 |
| Yahoo | 0.33210 | 0.34731 |

The parameters used were $m = 128^2$ (the number of poles or coefficients) and 13 for the sliding data window in the historical volatility estimate. The length of the forecast vector was set to 64. This was established by means of the *Maximum Entropy* spectrum used to extract a basic cycle in the squared log returns (conditional variance). The value of *m* was found by way of the following relationship:

$$m = \frac{cycle\ length}{h}$$

---

[2] LPC uses more coefficients in order to allow a lower information bit rate.

Where *h* = the Hurst exponent[3], estimated at 0.5.

The sliding data window value was established by means of the best fit EWMA parameter: $\lambda = 0.928571429$ of the peaks of the Maximum Entropy spectrum.

The volatilities projected with LPC were forecast in *t* + 1 for the purpose of comparison. The fact that the values obtained are so close is a clear sign that the processes are essentially very similar; however, LPC can yield a longer term projection, which offers a more accurate estimate (i.e. closer to the observed implied volatility) than that afforded by GARCH. It also suggests a method for obtaining an estimate of the structural volatility of a signal when used in conjunction with the Hurst exponent[4]. The notion that the process can be explained by fBm tells us that the Hurst exponent, and consequently the fractal dimension are important metrics in the determination of the signal's underlying non linearity.

As pointed out above, this algorithm also performs significantly faster than GARCH given its relative computational ease.

---

[3] See Appendix I
[4] See Appendix II

# Appendix I

# The Hurst Exponent and the Probability Density Function

Consider a self-similar function $y(t)$. The difference between the maximum and minimum values of $y$ in a time interval $\Delta t$ defines a range for that interval, $R(\Delta t)$. Given that $y$ is self-similar, the ensemble-averaged value of $R$ will scale with $\Delta t$. We can write:

$$\langle R(\Delta t) \rangle = c \Delta t^H \qquad (1.14)$$

where $c$ and $H$ are constants; $H$ defines the Hurst exponent. For data that are only approximately self-similar, we use this relation to check their proximity to self-similarity, and also to obtain an effective value for $H$. We proceed as follows: create a moving window $\Delta t$ one point at a time through the raw data; an array of values $R(\Delta t)$ is created from which the mean $\langle R \rangle$ is found, thus reducing the effects of uneven sampling. This is repeated for a range of $\Delta t$ within the length of the data set. A plot of $\log \langle R(\Delta t) \rangle$ against $\log \Delta t$ will reveal any deviations from self-similarity, while the slope will yield the best estimate of $H$. Linear regression is utilized to calculate the 95% confidence interval for $H$.

Trivially, a function that is exactly constant over time has $H = 0$. At the other extreme, $H = 1$ indicates a function whose range increases linearly with time (for a positive $c$ in (1.14)). Intermediate values of $H$ are generated by fractal functions: Random

Gaussian noise possesses $H \approx 0.2$ while Gaussian Random Walks (whose next value in time is $y_{t-1} + \varepsilon$, where $\varepsilon$ is a random Gaussian increment) will yield $H \approx 0.5$. The value of $H$ does not uniquely establish correlation; however, uncorrelated series may present significant probabilities of observing greater values as the time-scale increases.

We now define $\alpha = \dfrac{1}{H}$ [5] as a dimension in the probability space defined for the characteristic function:

$$\log(f(t)) = i\sigma t - \gamma |t|^{\alpha}\left(1 + i\beta \frac{t}{|t|} \tan\left(\alpha \frac{\pi}{2}\right)\right) \quad (1.15)$$

where:

$t$ = a constant of integration.

$\sigma$ = the location parameter of the mean.

$\gamma$ = is the scale parameter to adjust differences in time frequency of data.

$\beta$ = is the measure of skewness with $\beta$ ranging between -1 and +1.

$\alpha$ = the kurtosis and the fatness of the tails. Only when $\alpha = 2$ does the distribution become equal to the Gaussian distribution.

The growth of range method for measurement of $H$ may be somewhat insensitive as a measure of correlation. In principle, we can define a measure of correlation relative to $H$ as:

$$\beta = 2H + 1 \quad (1.16)$$

---

[5] Peters, Edgar "Chaos and Order in the Capital Markets", New York: J. Wiley & Sons (1995)

following Malamud & Turcotte. We would then have $\beta = 0$ for uncorrelated Gaussian noise and $\beta = 2$ for a Gaussian Random Walk. For $\beta$ between 1 and 2 or $H$ between 0 and 0.5 we have a mean reverting process.

Observe the values below. They cross asset type and class:

| Asset | P | H | α | β |
|---|---|---|---|---|
| Bond 10 Yr. | 1.29126 | 0.441001 | 2.26757 | 1.88200 |
| Bovespa | 1.07707 | 0.596789 | 1.67563 | 2.19358 |
| Dow | 0.34356 | 0.515913 | 1.93831 | 2.03183 |
| Google | 1.07864 | 0.501961 | 1.99218 | 2.00392 |
| MSFT | 0.24719 | 0.50076 | 1.99696 | 2.00152 |
| Nikkei | 1.90751 | 0.448339 | 2.23046 | 1.89668 |
| S&P | 0.44598 | 0.529237 | 1.88951 | 2.05847 |
| Yahoo | 0.33822 | 0.475057 | 2.10501 | 1.95011 |
| Average | | 0.501132 | 2.01196 | 2.00226 |

It is clear that the $H$ values are consistent across widely varied asset classes and even national boundaries, while $P$ values (maximum likelihood estimates of a Pareto Distribution shape parameter) are not. This tells us that $H$, as a gauge of long and short-term memories, and $\beta$, as a derived correlation metric, measure the same underlying system.

Noteworthy is the fact that the values for $\alpha$ are quite close to 2. This tells us that the processes are quasi Gaussian, hence justifying the use of the LPC model when estimating volatility.

# Appendix II

## Proof of the Relationship between $D_B$ and $H$:

$$D_B = 2 - H \quad (1.17)$$

**Proof:** Recall that to calculate the box counting dimension $D_B$ of a fractal object, we cover the object with $N$ boxes of side length $\delta$ and then compute $D_B$ using:

$$D_B = \frac{\log(N) - \log(V^*)}{\log\left(\frac{1}{\delta}\right)} \quad (1.18)$$

Where $N$ = the number of boxes of length $\delta$ used to cover a line segment and $\frac{V^*}{\delta}$ is the minimum count of one dimensional boxes needed to cover said line segment.

For a fractional Brownian motion trace, suppose that we isolate a time series of $T$ time steps that spans 1 unit of time. During each time step of length $\frac{1}{T}$ the average vertical range of the function is $\left(\frac{1}{T}\right)^H$ due to the scaling properties of any self-similar process. In order to cover the plot of the function during a single time step, a rectangle of width $\frac{1}{T}$ and height $\left(\frac{1}{T}\right)^H$ is required.

The area of this rectangle is $\left(\frac{1}{T}\right)^{H+1}$, so the number of squares with side length $\frac{1}{T}$ needed to cover it is $\left(\frac{1}{T}\right)^{H-1}$. For all $T$ of the time steps, the total number of squares needed to cover the plot of the function is $\left(\frac{1}{T}\right)^{H-2}$. If we let $N = \left(\frac{1}{T}\right)^{H-2}$ and $\delta = \frac{1}{T}$, then the box counting dimension is given by:

$$D_B = \frac{\log\left(\frac{1}{T}\right)^{H-2}}{\log(T)} = 2 - H \quad (1.19)$$

The time series spans 1 unit of time, so $V^* = 1^{D_B} = 1$.

This relationship between the fractal dimension and the Hurst exponent aligns perfectly with the notion of fractal dimension as a measure of the roughness of an object. As $H$ increases and the fractional Brownian motion displays greater persistence, the plot of the function becomes smoother and $D_B$ decreases accordingly. Conversely, as $H$ decreases and the fractional Brownian motion is more anti-persistent, the plot of the function becomes more jagged and $D_B$ increases.

The literature demonstrates the existence of a parametric function in which each coordinate's function is a Brownian motion trace. Similarly, we can construct fractional Brownian motion paths from traces. To calculate the box counting dimension of a fractional Brownian motion path with two coordinates, we examine a section of the path that results from $T$ time steps spanning 1 unit of time. During each time step of length $\frac{1}{T}$, each of the two traces has range $\left(\frac{1}{T}\right)^H$, so we can cover the path during a single time step with a square of side length $\left(\frac{1}{T}\right)^H$. For all $T$ time steps, the path requires $T$ such squares to cover it. Letting $N = T$ and $\delta = \left(\frac{1}{T}\right)^H$, we have:

$$D_B = \frac{\log(T)}{\log\left(\left(\frac{1}{T}\right)^H\right)} = \frac{1}{H} \quad (1.20)$$

Since $H$ can fall between 0 and 1, $\frac{1}{H}$ can assume values greater than 2. However, the fractal dimension of the path cannot exceed its Euclidean dimension $D_E$, so we modify the box counting dimension to be:

$$D_B = \min\left(\frac{1}{H}, D_E\right) \quad (1.21)$$

Thus, the fractal dimension of regular Brownian motion and anti-persistent fractional Brownian motion paths is 2, and the fractal dimension of persistent paths lies between 1 and 2.